\documentstyle[aps,prl,twocolumn,epsf]{revtex}

\begin{document}
\draft

\newcommand{\bomega}{\mbox{\boldmath{$\omega$}}}

\twocolumn[\hsize\textwidth\columnwidth\hsize\csname@twocolumnfalse%
\endcsname

\title{Gravity-driven Dense Granular Flows}
\author{Deniz Erta{\c s}$^1$\cite{*}, Gary S. Grest$^2$, Thomas C. Halsey$^1$, 
Dov Levine$^3$, 
and Leonardo E. Silbert$^2$}
\address{$^1$ Corporate Strategic Research, ExxonMobil Research and Engineering, 
Route 22 East, Annandale, New Jersey 08801\\
$^2$ Sandia National Laboratories, Albuquerque, New Mexico 87185-1411\\
$^3$ Department of Physics, Technion, Haifa, 32000 Israel}
\date{\today}
\maketitle

\begin{abstract}
We report and analyze the results of numerical studies of dense granular 
flows in two and three dimensions, using both linear damped springs and
Hertzian force laws between particles. Chute flow generically produces 
a constant density profile that satisfies scaling relations suggestive 
of a Bagnold grain inertia regime. The type of force law has little 
impact on the behavior of the system. Bulk and surface flows differ
in their failure criteria and flow rheology, as evidenced by the change 
in principal stress directions near the surface. Surface-only flows are 
not observed in this geometry.
\end{abstract}
]

Understanding the behavior of granular materials has been a great challenge
to scientists\cite{Coulomb,Behringer} and engineers\cite{Nedderman,Brown}. 
One major hurdle has been the lack of a formal connection between the 
complicated but relatively well-understood world of contact 
mechanics\cite{Johnson}, which describes the nature and dynamics of 
intergranular interactions, and empirical continuum models 
that describe the macroscopic behavior of the system. 

\begin{figure}[b]
\centerline{\epsfxsize=3.1in \epsffile{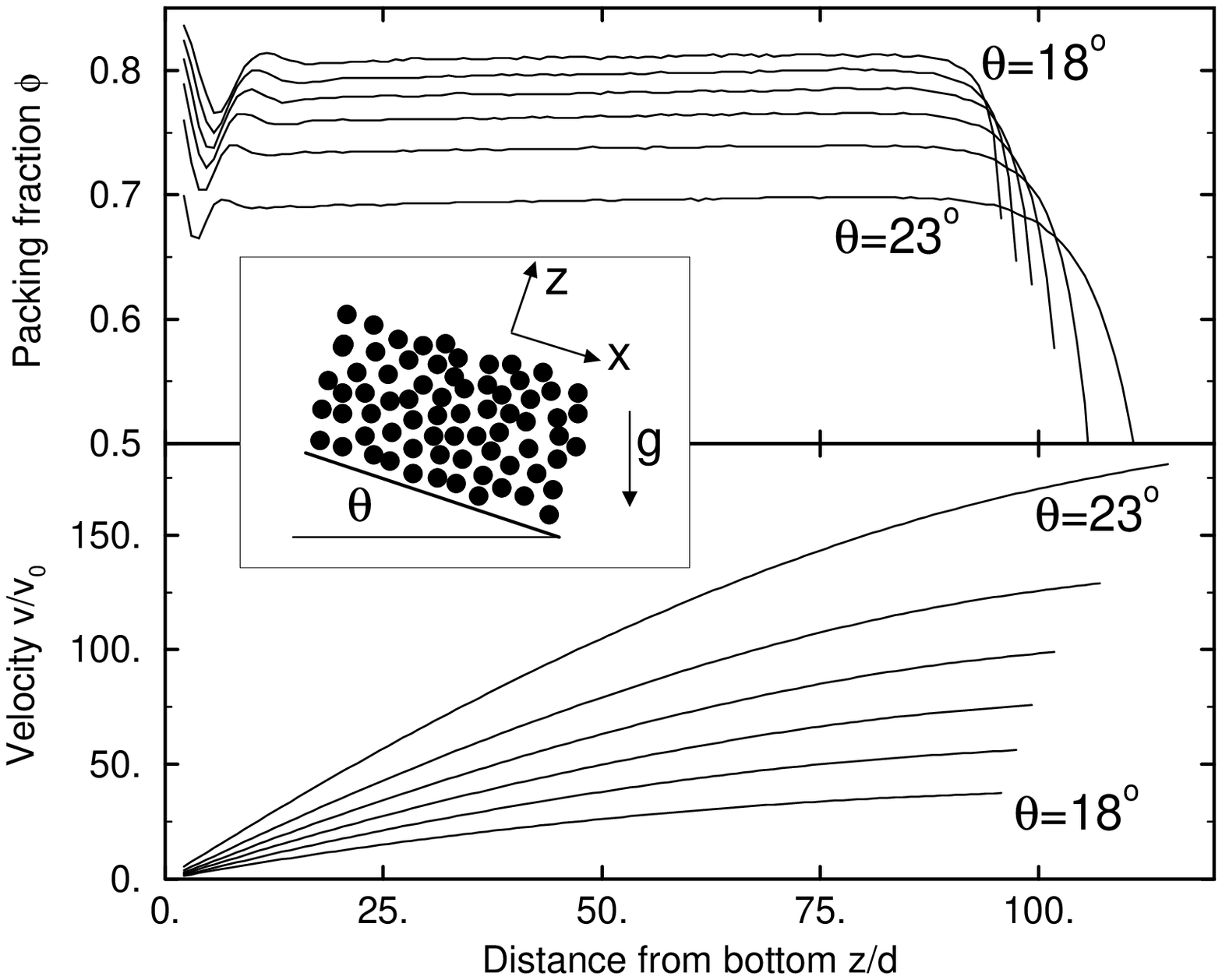}}
\centerline{\epsfxsize=3.1in \epsffile{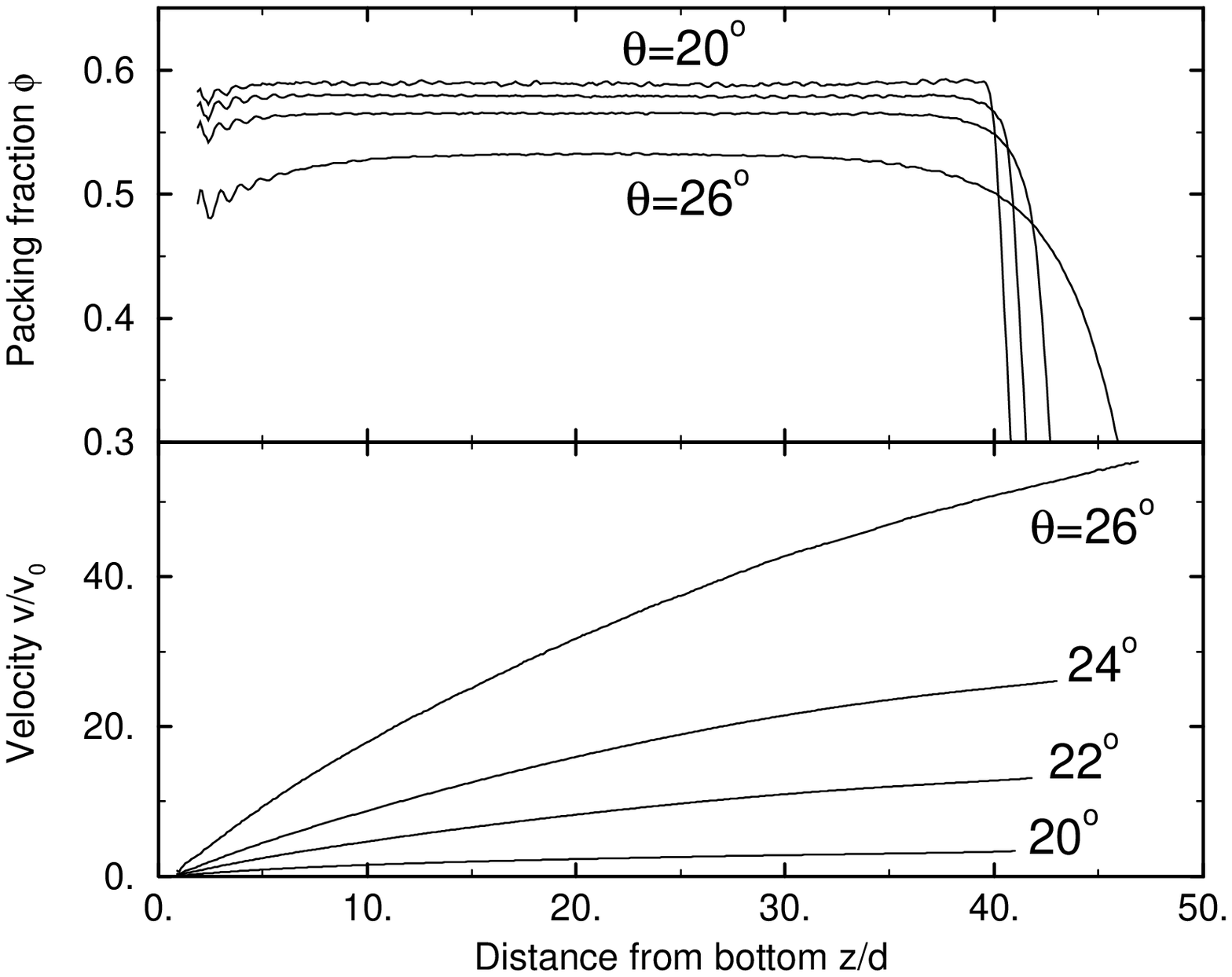}}
\caption{Density and velocity profiles as a function 
of height from the bottom of the assemblies for 2D (top) and 
3D (bottom) simulations. Inset shows a schematic of the geometry.
Results are for $10\, 000$ particles in 2D and $8\, 000$ in
3D. The characteristic velocity $v_0=\sqrt{gd}$.}
\label{figprofiles}
\end{figure}

We aim to isolate the essential features of granular flow, unencumbered
by complicated boundary effects.  To this end, we perform simulations
of granular dynamics in a simple geometry: gravity driven
dense flow down an inclined plane, denoted henceforth as ``chute flow".
Several remarkable features emerge:

(1) In steady-state, the packing fraction $\phi$ remains constant 
as a function of depth, beyond a dilatant surface region a few 
layers thick (Fig.~\ref{figprofiles}.) The compacting influence of 
increasing stress due to the weight of grains overhead is 
balanced by increasing velocity fluctuations towards the bottom of the 
assembly.

(2) Unlike Couette flows, the entire assembly is in motion and 
surface-only flows are not observed. 

(3) Components of the stress tensor and the square of the strain
rate grow linearly with depth, indicative of Bagnold grain-inertia
behavior\cite{Bagnold}.

\begin{figure}
\centerline{\epsfxsize=3.1in \epsffile{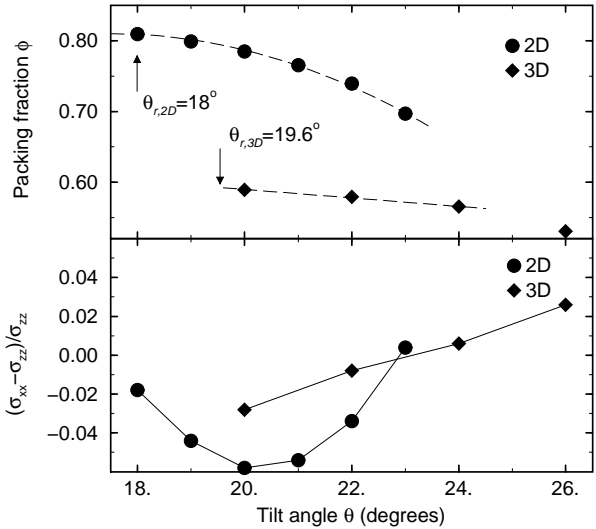}}
\caption{Tilt dependence of the packing fraction (top) and the normal
stress anomaly $(\sigma_{xx}-\sigma_{zz})/\sigma_{zz}$ (bottom) below 
the transitional surface layer.
The dashed lines are fits to the forms Eq.~(\protect\ref{eqthetadepa}
-\protect\ref{eqthetadepb}). The solid lines are guides to the eye.
}
\label{figthetadep}
\end{figure}

(4) Normal stresses differ from each other\cite{Savage79} 
in a systematic way (Fig.~\ref{figthetadep}), which we do not 
fully understand.
  
We report results of large scale molecular dynamics simulations of 
chute flow in two and three dimensions (2D and 3D), with interparticle 
interactions betweeen the (monodisperse) spheres modeled using 
both damped linear springs and Mindlin-Hertz contact forces, with static 
friction. Detailed results of the simulations will be presented 
elsewhere\cite{longpaper}. 
The main obstacle for experiments and simulations so far has been 
the difficulty of reaching and maintaining steady state. Previous 
simulations\cite{Walton} employed 
very few particles or did not reach steady state\cite{Poschel}. 
All of these simulations were in 2D with the exception of 
Walton\cite{Walton}. Experiments on chute flow\cite{Drake,Pouliquen99}
did not involve deep assemblies. Different effects of flow were also 
studied in simulation, such as size segregation\cite{Hirshfield}. 

The 3D simulation cells contain spheres of diameter $d$ and mass
$m$, supported by a fixed bottom on the $x-y$ plane. The bottom wall
is constructed from a cross-section of a random close packing of 
identical spheres, providing a rough surface. Periodic boundary 
conditions are imposed along $x$ and $y$ directions. 
2D simulations follow the same procedure, except that particles
are restricted to the $x-z$ plane and the bottom consists of
a regular array of particles of diameter $2d$. In both cases,
there is no slip observed at the bottom. (There is some slip in 2D  
with a regular array of particles of diameter $d$ at the bottom.)
The gravity vector {\bf g} is rotated  by the tilt angle 
$\theta$  away from the $(-\hat z)$ direction in the $x-z$ plane, 
so that the free surface is normal to the $\hat z$ axis. In 3D, 
most of our results are for $8\, 000$ particle
systems with a simulation cell of size $L_x=18.6d$ and 
$L_y=9.3d$, resulting in an assembly roughly 40 particles deep 
at rest. In 2D,  $L_x=100d$ and the number of particles varied from 
a few hundred to $20\, 000$. In this Letter, we present
results for $N=10\, 000$, with a depth of about 100 particles. 

We use the contact force model of Cundall and Strack\cite{Cundall}.
Static friction is implemented by keeping track of the elastic
shear displacement throughout the lifetime of a contact.
For two contacting particles at 
positions {\bf r}$_1$ and {\bf r}$_2$, with velocities {\bf v}$_{1,2}$ 
and angular velocities $\bomega_{1,2}$, the force on particle 1 is computed 
as follows: The normal compression $\delta$, normal velocity ${\bf v}_n$,
relative surface velocity ${\bf v}_t$, and the rate of change of the 
elastic tangential displacement ${\bf u}_t$, set to {\bf 0} at the 
initiation of a contact, are given by
\begin{eqnarray}
\delta&=&d-||{\bf r}_{12}||, \\
{\bf v}_n&=&({\bf v}_{12}\cdot {\hat r}_{12}){\hat r}_{12}, \\
{\bf v}_t&=&{\bf v}_{12}-{\bf v}_n-(\bomega_1+\bomega_2)
\times{\bf r}_{12}/2, \\
\frac{d{\bf u}_t}{dt}&=&{\bf v}_t-\frac{({\bf u}_t\cdot{\bf v}_{12})
{\bf r}_{12}}{||{\bf r}_{12}||^2},
\label{equt}
\end{eqnarray}
where ${\bf r}_{12}={\bf r}_1-{\bf r}_2$, ${\hat r}_{12}
={\bf r}_{12}/||{\bf r}_{12}||$, and ${\bf v}_{12}={\bf v}_1-{\bf v}_2$. 
The second term in Eq.(\ref{equt}) arises from the rigid body rotation 
around the contact point and assures that ${\bf u}_t$ always remains
in the local tangent plane of contact. Normal and tangential forces 
acting on particle 1 are given by
\begin{eqnarray}
{\bf F}_n&=&k_n f(\delta/d)\left(\delta{\hat r}_{12}-\tau_n{\bf v}_n\right), \\
{\bf F}_t&=&k_s f(\delta/d)\left(-{\bf u}_t-\tau_s{\bf v}_t\right),
\end{eqnarray}
where $k_{n,s}$ and $\tau_{n,s}$ are elastic constants and viscoelastic
relaxation times respectively; $f(x)=1$ for damped linear springs
or $f(x)=\sqrt{x}$ for Hertzian contacts between spheres.
A local Coulomb yield criterion, $F_t<\mu F_n$, is satisfied by
truncating the magnitude of ${\rm \bf u}_t$ as necessary.
Thus, the contact surfaces are treated as ``stuck" while $F_t<\mu F_n$,
and slipping while the yield criterion is satisfied. This ``proportional
loading" approximation\cite{groove} is a simplification of the much more 
complicated and hysteretic behavior of real contacts\cite{Mindlin}. 
The force on particle 2 is determined from Newton's third law.
Each particle is also subject to a body force
\begin{equation}
{\bf F}_{\rm body}=mg(-{\hat z}\cos\theta+{\hat x}\sin\theta).
\end{equation}
All results are given in terms of non-dimensionalized 
quantities: Distances, times, velocities, forces, elastic 
constants and stresses are reported in units of $d$, 
$t_0\equiv\sqrt{d/g}$, $v_0\equiv\sqrt{gd}$, $F_0\equiv mg$,
$k_0\equiv mg/d$ and $\sigma_0\equiv mg/d^2$, respectively. 
The summary of parameters used in the simulations are shown in 
Table~\ref{tabparam}. In these units, the correct elastic constant 
for glass spheres with $d=100\mu m$ would have been
$k_n^{\rm glass}/k_0\approx 3\times 10^{10}$, which would have 
prohibited a large-scale simulation. We use $k_n/k_0=2\times 10^5$, 
while controlling the coefficient of restitution for binary 
collisions through $\tau_n$, assuming that we remain sufficiently
close to the $k_n\to\infty$ limit of small deformations. 
Simulations for $k_n/k_0=2\times 10^4$ and
$2\times 10^6$ in 2D  gave essentially the same results,
supporting this assumption.
For Hertzian contacts, the ratio $k_s/k_n$ depends on the Poisson 
ratio of the material\cite{Mindlin}, and is about $2/3$ for most 
materials. We use the value $k_s/k_n=2/7$, since this makes the
period of normal and shear contact oscillations equal to each other 
in the damped linear springs case\cite{Schafer}. The collisional 
dynamics are not very sensitive to the precise value of this ratio. 
\begin{table}
\begin{tabular}{ccccccc}
Dimen-&Force&$k_n/k_0$&$\tau_n/t_0$&$k_s/k_n$&$\tau_s/\tau_n$&$\mu$\\
sion&Law&\\
\hline
2D&Linear&$2\times 10^5$&$8.375\times10^{-5}$&2/7&0&0.5\\
3D&Hertzian&$2\times 10^5$&$1.25\times10^{-4}$&2/7&0&0.5
\end{tabular}
\caption{Simulation parameters for results shown.
The normal coefficient of restitution for 2D simulations 
is 0.92; it is a function of collision velocity for Hertzian 
contacts in 3D simulations.}
\label{tabparam}
\end{table}

The equations of motion for the translational
and rotational degrees of freedom  were integrated with either
a third-order Gear predictor-corrector or velocity-Verlet scheme with 
a time step $\delta t=1\times 10^{-4}$. Typically
it was necessary to run between 5 and $20\times 10^6~\delta t$
to reach steady state, particularly when starting from a non-flowing
state. All coarse grained quantities have been averaged both temporally 
(typically 2 to $8\times 10^6~\delta t)$ and spatially over slices 
of constant $z$.  

The main characteristics of all the flows are: (i) The existence
of an ``angle of repose" $\theta_r$, such that granular flows can not
be sustained for $\theta<\theta_r$, (ii) a steady-state flow with a 
packing fraction independent of depth
for $\theta_r<\theta<\theta_{\rm max}$, and
(iii) for $\theta >\theta_{\rm max}$, development of a shear thinning layer 
at the bottom of the assembly that results in lift-off and unstable acceleration 
of the entire assembly. 
For very thin assemblies, less than about  $20$ layers, the value of 
$\theta_r$ depends on their depth, in agreement with 
experiment\cite{Pouliquen99}. Here we consider only deep assemblies where 
the value of $\theta_r$ is independent of depth, and  focus our attention 
on  region (ii).
As seen in Fig.~\ref{figprofiles}, the packing fraction $\phi$ 
remains constant as a function of depth, away from the free surface 
and the bottom wall. Its value is shown as a function of $\theta$ in 
Fig.~\ref{figthetadep}. Results in 2D for systems of size $5\, 000$ 
and $20\, 000$ demonstrate that the thicknesses of the boundary 
layers at the bottom and top are independent of the height of the
assembly. The  data suggest an approximate tilt dependence for $\phi$ of  
the form
\begin{eqnarray}
\phi_{2D}(\theta)&\approx&
\phi_{2D}^{\rm max}-c_{2D}(\theta-\theta_{r,2D})^2, 
\label{eqthetadepa} \\
\phi_{3D}(\theta)&\approx&
\phi_{3D}^{\rm max}-c_{3D}(\theta-\theta_{r,3D}), 
\label{eqthetadepb}
\end{eqnarray}
where $\phi_{2D}^{\rm max}=0.810(5)$ and $\phi_{3D}^{\rm max}=0.595(5)$.

\begin{figure}
\centerline{\epsfxsize=3.1in \epsffile{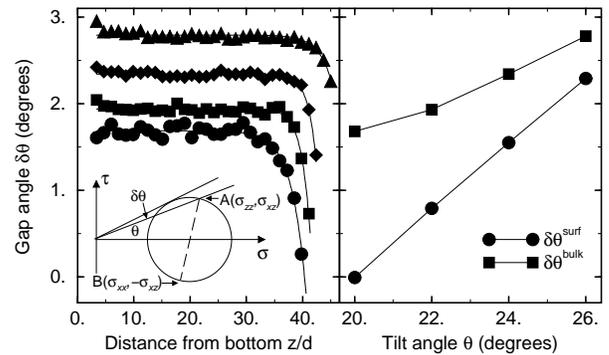}}
\caption{{\it Left:} The gap angle $\delta\theta$
(defined pictorially in the inset) as a function of height 
for $\theta=20^\circ$ (circles), $22^\circ$ 
(squares), $24^\circ$ (diamonds), and $26^\circ$ (triangles). The
lines are fits that decay exponentially from 
$\delta\theta^{\rm surf}$ to $\delta\theta^{\rm bulk}$ with a typical 
decay length of 1.5 to $2.5d$, inidcating a surface layer about $5d$ to 
$8d$ thick. {\it Right: } 
Although $\delta\theta^{\rm surf}$
vanishes at $\theta=\theta_r$, $\delta\theta^{\rm bulk}$ remains finite,
suggesting that the initiation of flow is controlled by the surface rather 
than the bulk (see text.)}
\label{figMohr}
\end{figure}

In 2D, upon lowering the tilt angle below $\theta_r$, we observe a 
compaction to a polycrystalline triangular lattice
with $\phi_{2D}\approx0.9$. This causes considerable hysteresis in 
2D simulations as $\theta$ is subsequently increased beyond $\theta_r$: 
There is no flow until a maximum angle of stability is exceeded.
Initial failure always occurs at the bottom, followed by movement of 
a dilation front towards the top. Once the system reaches
steady state, $\theta$ can be reduced and the system continues
to flow while $\theta >\theta_r$. 
On the other hand, in 3D there is no jump in $\phi$ as the system 
comes to a stop, and no detectable hysteresis as the system is stopped and
restarted by varying $\theta$. Thus, Mohr-Coulomb analysis of the stress 
tensor\cite{Nedderman} can be used to relate the flow rheology near 
$\theta_r$ to the Coulomb failure criterion as follows:
The Mohr circle is the set of normal and shear stresses 
($\sigma$ and $\tau$, see inset in Fig.~\ref{figMohr}) associated
with all possible shear planes. The points $A$ and $B$ at coordinates 
$(\sigma_{zz},\sigma_{xz})$ and $(\sigma_{xx},-\sigma_{xz})$ in the
$(\sigma,\tau)$ plane form a diameter of the circle. 
At a given tilt angle, $\sigma_{zz}$ and $\sigma_{xz}$ are determined 
by force balance, which pins down the location of point $A$ 
($\sigma_{xz}/\sigma_{zz}=\tan\theta$; see Fig.~\ref{figMohr}, inset.) 

\noindent However, $\sigma_{xx}$, thus the location of point $B$, remains 
indeterminate from these considerations. The ``gap angle" 
$\delta\theta\equiv\theta_{MC}-\theta$ is a measure of the difference 
between $\tau/\sigma$ along the slip plane parallel to the surface 
($=\tan\theta$) and the largest value of $\tau/\sigma$ experienced 
by the system (equal to the slope $\tan\theta_{MC}$ of the line 
that passes through the origin and that is tangent to the Mohr circle.) 
For an ideal Coulomb material with a uniform yield criterion, 
$\delta\theta=0$ at $\theta=\theta_r$, when the static system is 
at incipient yield. The behavior of $\delta\theta$ as a function of
depth at different tilt angles, as shown in Figure~\ref{figMohr},
reveal that: (i) $\delta\theta$ becomes independent of depth 
below a transitional surface layer about $5d$ to $8d$ in thickness, 
(ii) the gap angle remains finite at the bottom and in the bulk 
but vanishes at the top surface when $\theta=\theta_r$. It thus seems 
that although the bulk of the system has the intrinsic capability 
to withstand slightly larger tilt angles, the destabilization of 
the surface at $\theta=\theta_r$ is ultimately responsible for 
the failure and initiation of flow in the entire system. Note that 
the transitional surface layer is not directly related to 
the dilatant layer; it is much thicker near $\theta=\theta_r$ 
and penetrates well into the region of constant density. 
Interestingly, in 2D the gap angle at $\theta_r$ is actually 
larger at the surface compared to the bulk, and both gap angles 
remain finite. However, the presence of hysteresis precludes us 
from applying the Mohr-Coulomb analysis in 2D.

A feature that distinguishes granular flows from Newtonian fluid 
flows is that normal stresses, i.e., diagonal terms in the stress 
tensor, are in general not equal to each other\cite{Savage79}.
Although $\sigma_{xx}\approx\sigma_{zz}$ in our simulations, 
we observe small but systematic deviations, which are depicted
in Fig.~\ref{figthetadep} by plotting the normal stress 
anomaly in the bulk, $(\sigma_{xx}-\sigma_{zz})/\sigma_{zz}$, as
a function of tilt angle. These deviations are likely to be due to 
a constitutive equation of the flow rheology which we have not 
yet been able to determine. In addition, in all 3D runs, 
$\sigma_{yy}$ is smaller than the other normal stresses by 
$10-15\%$, suggesting that consolidation and compaction normal to 
the shear plane is poorer.
 
\begin{figure}
\centerline{\epsfxsize=3.1in \epsffile{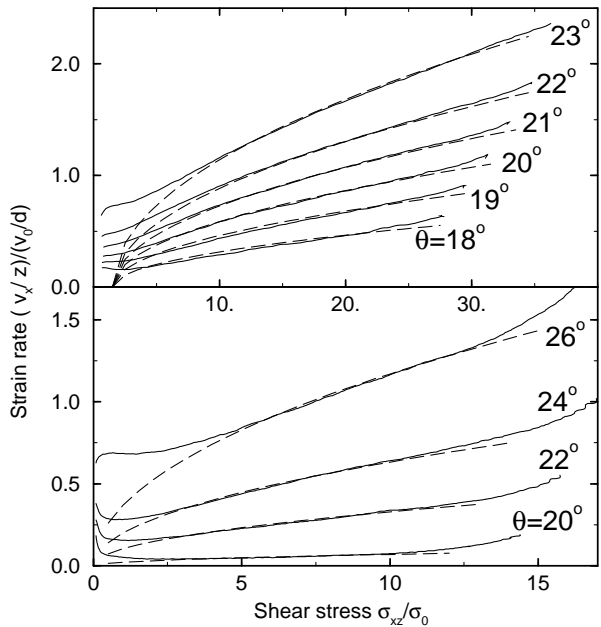}}
\caption{Strain rates plotted as a function of shear stress $\sigma_{xz}$
for 2D (top) and 3D (bottom) reveal Bagnold scaling 
(dashed lines) in the constant packing fraction regime 
away from the free surface and the bottom wall. For 2D, an offset of
$1.5\sigma_0$ in the shear stress was needed for an acceptable fit.}
\label{figstrainrate}
\end{figure}

Another question of particular interest is the relationship between
the stresses and strain rate. Shear stress $\sigma_{xz}$ is a 
linear function of strain rate $\dot\gamma\equiv\partial v_x/\partial z$ 
for viscous fluids, and quadratic in $\dot\gamma$ for granular systems 
in the Bagnold grain-inertia regime. The latter result is rather
general: When typical stresses on grains are large compared to the weight 
of individual particles but not large enough to significantly distort 
the spheres $(1 \ll \sigma/\sigma_0 \ll k_n/k_0)$, 
the only relevant time scale is $\dot\gamma^{-1}$, which forces
$\sigma_{xz}\propto\dot\gamma^2$ simply by dimensional analysis. As an example,
Fig.~\ref{figstrainrate} shows the relationship between shear strain 
rate $\dot\gamma$ and shear stress $\sigma_{xz}$ for 2D and 3D cases. 
Below the transitional surface layer, 
and away from the bottom wall, both systems exhibit Bagnold scaling, 
indicated by the dashed lines in Fig.~\ref{figstrainrate}. Data for 2D 
suggest an offset of about $1.5\sigma_0$ in the stress, possibly due to 
corrections from the body force on individual spheres. Such an offset 
is not needed for an acceptable fit to the 3D data.

In order to probe the sensitivity of the results to the exact form of 
the stiff elastic response, we have also performed runs in the 3D system
with linear damped springs, keeping all other parameters 
fixed\cite{longpaper}.
Remarkably, the packing fraction profiles and the normal stress
anomaly remained virtually the same, and strain rate profiles were
changed only by a global factor of about 1.35, suggesting that 
results are not too sensitive to the particular force scheme selected.

The lack of a regime with only surface flow is in contrast with experimental 
observations of avalanche flows in rotating drums\cite{Behringer}. 
Since periodic boundary conditions are used, our simulations correspond 
to an infinitely large system with finite depth and fixed surface tilt, 
and therefore constitutes a different system. Experiments on chute 
flows\cite{Savage79} also lack a regime of surface only flow; although
this might be attributed to side-wall friction that necessitates 
higher tilt angles to initiate flow. Although there is no 
fundamental reason we can find that prohibits surface-only flows in 
this geometry, it appears that they are unlikely to occur in an 
assembly of monodisperse spheres.

DL is supported by the Israel Science Foundation
under grant 211/97. Sandia is a multiprogram laboratory 
operated by Sandia Corporation, a Lockheed 
Martin Company, for the United States Department of Energy 
under Contract DE-AC04-94AL85000.


\end{document}